\documentclass[]{raa}            
\usepackage{graphicx,times}
\usepackage{amssymb}
\usepackage{natbib}
\begin{document}
   \title{LAMOST J052016.79+345651.7: An EW-type Binary with Emission Line Spectra and Circumstellar Material}
\volnopage{ {\bf 0000} Vol.\ {\bf 0} No. {\bf XX}, 000--000}
\setcounter{page}{1}
\author{Y. H. Chen\inst{1}$^*$, \ C. M. Duan\inst{1,2}, \ and Y. Yu\inst{1}}
\institute{\inst{1} Institute of Astrophysics, Chuxiong Normal University, Chuxiong 675000, China; {$yanhuichen1987@126.com$}\\
           \inst{2} Faculty of Science, Kunming University of Science and Technology, Kunming 650093, China\\
\vs \no
{\small Received [0000] [July] [day]; accepted [0000] [month] [day] }}

\abstract{LAMOST J052016.79+345651.7 was identified as an EW-type eclipsing binary by Chen et al. when studying the periodic variable stars based on the ZTF telescope. An orbital period of 0.3507818 days has been reported. Using the ZTF g, r, i band light curves, we reproduced the orbital period and obtained a phase folded diagram. The multi-band apparent magnitudes from Pan-STARRS, 2MASS, and WISE, the color indices, and the infrared excess in the WISE w4 band all indicate that LAMOST J052016.79+345651.7 contains cold circumstellar material. All 19 low resolution spectra from LAMOST exhibit prominent H$\alpha$ emission lines, along with clear N II and S II emission lines, indicating that LAMOST J052016.79+345651.7 contains optically thin, warm ionized gas with extremely low electron density. From the 19 spectra, an effective temperature of 6200 $\pm$ 400\,K can be obtained. Therefore, the emission lines likely originate from shock-driven outflows and mass ejection, while the infrared excess likely comes from dust formed as the outflows cool. We performed light curve fitting and evolutionary simulation study on LAMOST J052016.79+345651.7 using PHOEBE and MESA, respectively. The phase shift of the fitted model can be uniquely determined, but the inclination and the $fillout_{factor}$ are degenerate. Considering the angular momentum loss process, the evolutionary simulation from detached binaries to contact binaries can reflect the main evolutionary process of LAMOST J052016.79+345651.7. When the orbital period of the evolutionary model is 0.351 days, the other parameters are generally consistent with the observed characteristics of LAMOST J052016.79+345651.7. Considering mass loss, LAMOST J052016.79+345651.7 will evolve into a blue straggler star in the future. Future LAMOST medium resolution time-domain spectra of LAMOST J052016.79+345651.7 may offer an opportunity to reveal more detailed physical processes.
 }
\authorrunning{Y. H. Chen, C. M. Duan, \& Y. Yu}       
\titlerunning{An EW-type Binary with Emission Line Spectra}  
\maketitle

\section{Introduction}

Stars are the dominant component of celestial bodies. Duch$\hat{e}$ne and Kraus (2013) provided a comprehensive and in-depth review of stellar multiplicity. They reported that the multiplicity rate and the breadth of orbital period distribution increase steeply with primary mass among field objects, and that the mass ratio distribution remains essentially flat for most populations other than the lowest mass objects. Even for a sample of low-mass, solar-type stars, the fraction of binary systems is above 33\% (Raghavan et al. 2010). Physical binaries are systems in which two stars are gravitationally bound to each other and orbit their common center of mass. Through photometric observations, spectroscopic observations, and high-precision astrometry, physical binaries can be classified as eclipsing binaries, spectroscopic binaries, and astrometric binaries, respectively. Cataclysmic variables (CVs) and X-ray binaries are ideal objects for studying accretion physics. Binaries contain a wealth of physical laws, and studying them is of general significance.

The Zwicky Transient Facility (ZTF, Bellm et al. 2019) is a new optical time-domain survey that uses the 48-inch Schmidt telescope and features a wide-field camera with a 47 deg$^{2}$ field of view and 8\,seconds readout time. Based on ZTF DR2 data reaching down to an r-band magnitude of 20.6, Chen et al. (2020) newly discovered or newly classified 621,702 periodic variables, including $\sim$ 350,000 eclipsing binaries. As early as 1978, in the chapter 'Classes of Eclipsing Binaries and the Roche Lobe' of the book Double Stars, Heintz detailed that eclipsing binaries can be classified according to their eclipse characteristics into the EA Algol type (EA-type), the EB $\beta$ Lyrae type (EB-type), and the EW W Ursae Majoris type (EW-type). Tkachenko et al. (2012) used a computer program package that implements the New Algol Variable algorithm (Andronov 2012) to study the phenomenological parameters of these three types of eclipsing binaries and to quantify their classification. Large sky Area Multi-Object fiber Spectroscopic Telescope (LAMOST, Zhao et al. 2012) has a field of view of 20 square degrees, an effective aperture of 3.6-4.9 meters, and can simultaneously obtain spectra of 4,000 celestial objects in a single exposure. Using LAMOST observational data up to June 16, 2017, Qian et al. (2018) identified the spectral types of 3,196 EA-type binaries and studied the atmospheric parameters for 2,020 of them. Gaia (Gaia Collaboration et al. 2016) is a cornerstone mission in the science programme of the European Space Agency (ESA), performing astrometric observations to map the Milky Way in three dimensions with microarcsecond accuracy. Based on Gaia DR2, El-Badry \& Rix (2018) reported an extensive and pure catalogue of wide binaries within 200\,pc of the Sun, including more than 50,000 main sequence(MS)/MS binaries, more than 3,000 white dwarf(WD)/MS binaries, and nearly 400 WD/WD binaries. Mowlavi et al. (2023) published a catalog of eclipsing binary candidates containing 2,184,477 sources based on Gaia DR3, the largest catalogue to date in terms of number of sources, sky coverage, and magnitude range. Big data, multi-band, and multi-method approaches are effective ways to study binary physics.

Using three dimensional magnetohydrodynamic nested grid simulations, Machida et al. (2005) studied the evolution and fragmentation of a rotating magnetized cloud core and investigated the conditions for fragmentation and binary formation. Paxton et al. (2011) released a powerful stellar evolution code, Modules for Experiments in Stellar Astrophysics (MESA), which can simulate stellar evolution from the pre-MS to the WD stage, incorporating a rich set of physical processes. In 2015, Paxton et al. incorporated a binary module, which includes rich physical processes such as gravitational wave radiation, mass loss, spin-orbit coupling, magnetic braking, and mass transfer from Roche lobe overflow (RLOF). As early as 1971, Wilson and Devinney proposed a general procedure for computing monochromatic light curves of close eclipsing binary systems, taking into account rotational and tidal distortion, the reflection effect, limb darkening, and gravity darkening. The procedure has been continuously improved and refined over time, e.g., Wilson (1990), Van Hamme (2007), Wilson et al. (2010), and Wilson (2012), etc. Based on this procedure, Pr$\check{s}$a and Zwitter (2005) released an open source PHysics Of Eclipsing BinariEs (PHOEBE), which retains 100\% compatibility with the Wilson-Devinney procedure and incorporates observational spectra into the solution seeking process. In his book published in 2018, Pr$\check{s}$a elaborates in detail on the construction logic and core principles of the eclipsing binary modeling code PHOEBE. Simulation studies can explain the observed binary phenomena in detail.

Based on ZTF DR2 data, Chen et al. (2020) classified 781,602 periodic variables into 11 main types using an improved classification method. ZTF J052016.80+345651.8 was identified as an EW-type binary with an orbital period of 0.3507920\,d in g band and 0.3507818\,d in r band. In the SIMBAD astronomical database, apart from the statistical records of the eclipsing binary ZTF J052016.80+345651.8 by Chan et al. (2022) and Mowlavi et al. (2023), there are no other reference records of in-depth studies. It is worth noting that ZTF J052016.80+345651.8 is marked as LAMOST J052016.79+345651.7 in the LAMOST catalog (with a slight adjustment in coordinates). From October 9, 2023, to October 21, 2025, LAMOST J052016.79+345651.7 was observed 19 times with LAMOST low resolution search (LRS), and obvious emission line spectra were detected. Emission line spectra in EW-type binaries are uncommon, making this an interesting target source worthy of systematic study. In Sect. 2, we analyze the ZTF photometric data, reproducing and confirming its EW-type binary identity through the phase folded figure. At the same time, we analyze the multi-band apparent magnitude data of LAMOST J052016.79+345651.7. In Sect. 3, we analyze the LAMOST spectral data and conduct an in-depth study of the spectral emission lines. In Sect. 4, a preliminary fitting study of LAMOST J052016.79+345651.7 is performed using the PHOEBE procedure and the MESA program. In Sect. 5, we give a discussion and conclusions.

\section{Analysis of photometric data from ZTF}

Light curves provide the most direct evidence for eclipsing binaries. When the orbital period is short and the observational cadence is longer than orbital period, phase folding serves as a powerful corroboration of the periodicity. The Barbara A. Mikulski Archive for Space Telescopes (MAST, Hou et al. 2023) is a data archive that collects data released from telescopes including Hubble, Kepler, TESS, SDSS, GALEX, IUE, FUSE, etc. In the MAST data archive, no photometric or spectroscopic observation data are available for LAMOST J052016.79+345651.7. The light curve data from the ZTF telescope are all the publicly available light curve data for LAMOST J052016.79+345651.7.

\begin{figure}
\begin{center}
\includegraphics[width=11.5cm,angle=0]{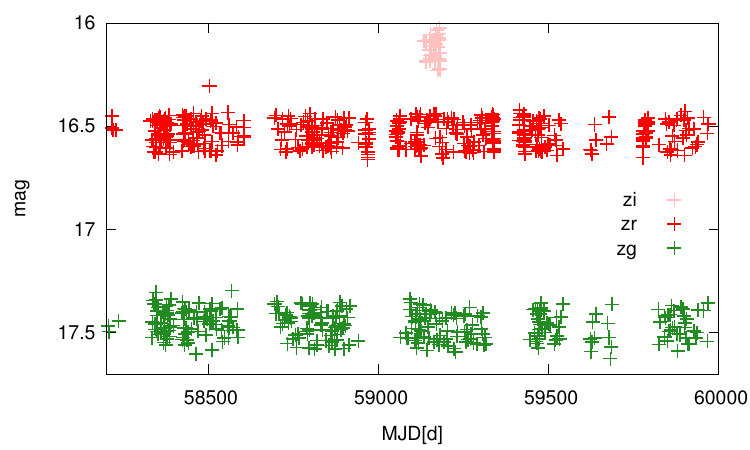}
\end{center}
\caption{Light curve figure for LAMOST J052016.79+345651.7 from ZTF. There are 337, 495, and 48 data points in the g, r, and i bands, respectively, for a cumulative total of 880 data points.}
\end{figure}

\begin{figure}
\begin{center}
\includegraphics[width=11.5cm,angle=0]{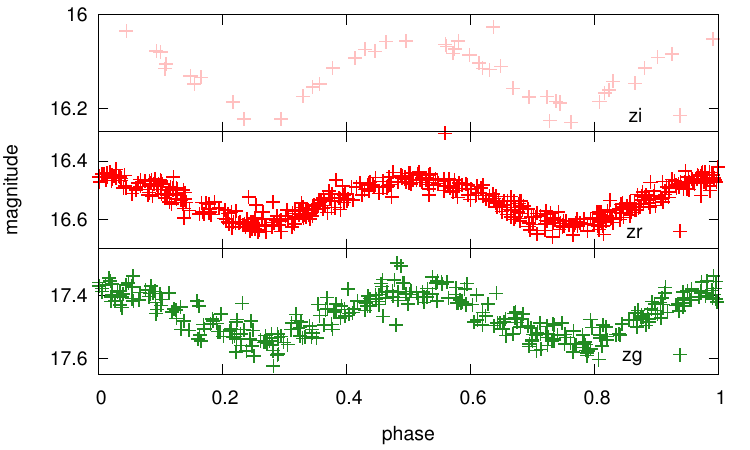}
\end{center}
\caption{Phase folded figure of the g, r, and i bands for LAMOST J052016.79+345651.7. The reference epoch (zero phase) is T0=58518.90245, and the folding period is P=0.350777\,d.}
\end{figure}

In Fig. 1, we show the light curve figure for LAMOST J052016.79+345651.7 from ZTF. The numbers of observations in the g, r, and i bands are 337, 495, and 48, respectively, and the durations are 1763.012\,d, 1755.150\,d, and 47.972\,d, respectively. The apparent magnitudes in these three bands are approximately 17.5, 16.5, and 16.1, respectively, with a magnitude dispersion of about 0.25. The intuitive variation trend of the light curve in Fig. 1 does not clearly reflect the periodic information of an eclipsing binary. Period04 (Lenz \& Breger 2005) is an open source software package designed for sophisticated time string analysis, especially dedicated to the statistical analysis of large astronomical time series containing gaps. We used the Period04 software to extract the strongest frequency signals from the ZTF g, r, and i band data, which are 5.70173001 d$^{-1}$, 5.70153649 d$^{-1}$, and 5.70025359 d$^{-1}$, respectively. It is worth noting that for the g band data, the large time intervals between data points impose a Nyquist frequency limitation. Therefore, we extracted the frequency using data points within the MJD range of 58770-58840 for the g band data, where the time intervals are smaller. Considering the orbital motion of the EW-type binary, the orbital periods of LAMOST J052016.79+345651.7 in the g, r, and i bands are 0.35077\,d, 0.35078\,d, and 0.35086\,d, respectively. We kept the first digit where they start to differ. These results are consistent with those reported by Chen et al. (2020), namely 0.3507920\,d in the g band and 0.3507818\,d in the r band. We took the average of the MJD values of the first observation point in the three bands, i.e., MJD=58518.90245, as the zero phase reference epoch (T0). By repeatedly fine tuning the orbital period (P) to fold the phases of the three bands data, we screened and fitted the orbital period based on the phase folded figure, and finally obtained an orbital period of 0.350777\,d. We tried using one additional significant digit. The phase folded figure of the three observation bands, as shown in Fig. 2, all indicate that LAMOST J052016.79+345651.7 is an EW-type eclipsing binary with an orbital period of 0.35777\,d.

\begin{table*}
\begin{center}
\tiny
\caption{Table of multi-band apparent magnitudes for LAMOST J052016.79+345651.7 from Pan-STRRS (Chambers et al. 2019), 2MASS (Cutri et al. 2003), and WISE (Cutri et al. 2012).}
\begin{tabular}{lcccccccccccccccccccccccccc}
\hline
filter             &g           &r          &i           &z            &y           &J           &H          &Ks        &w1        &w2        &w3         &w4        \\
$\lambda$($nm$)    &\,486.6     &\,621.5    &\,754.5     &\,867.9      &\,963.3     &\,1,240     &\,1,650    &\,2,160   &\,3,350   &\,4,600   &\,11,600   &\,22,100  \\
\hline
mag                &\,17.4635   &\,16.6785  &\,16.2292   &\,15.9839    &\,15.8586   &\,14.846    &\,14.329   &\,14.138  &\,14.026  &\,14.061  &\,13.543   &\,8.262   \\
err                &\,0.0066    &\,0.0140   &\,0.0141    &\,0.0100     &\,0.0086    &\,0.038     &\,0.044    &\,0.055   &\,0.045   &\,0.074   &           &\,0.356   \\
\hline
 color index       &            &g-r        &r-i         &i-z          &            &z-J         &J-H        &H-Ks      &          &          &           &          \\
\hline
mag                &            &0.785      &0.4493      &0.2453       &            &1.1379      &0.517      &0.191     &          &          &           &          \\
\hline
\end{tabular}
\end{center}
\end{table*}

In Table 1, we show a multi-band apparent magnitude table for LAMOST J052016.79+345651.7. The optical, near-infrared, and mid-infrared band data come from Pan-STRRS (Chambers et al. 2019), 2MASS (Cutri et al. 2003), and WISE (Cutri et al. 2012) respectively. We calculated some color index values for physical analysis of LAMOST J052016.79+345651.7. According to the color index values of z-J, J-H, and H-Ks in Table 1, by comparing with the synthetic photometry color index table (Covey et al. 2007), we find that the spectral type of LAMOST J052016.79+345651.7 appears to be closest to M0. However, the LAMOST LRS spectra of this source suggests that its spectral type is apparently A or F. This seems somewhat contradictory. We note that the w4 band magnitude is 8.262 $\pm$ 0.356 in Table 1, which implies that this source harbors cold circumstellar material. If the peak of the circumstellar material's blackbody radiation is near the w4 band, then the effective temperature ($T_{\rm eff}$) of the circumstellar material is approximately 131\,K. This resolves the contradiction: the color index values in Table 1 are the combined result of an A- or F-type star and cold circumstellar material at approximately 131\,K. The blue end color indices should be minimally affected by the circumstellar material. Indeed, the g-r value in Table 1 corresponds to a K2 (Covey et al. 2007) spectral type, which is closer to A or F than to M0. Taking into account the size of the errors in the apparent magnitude values, the apparent magnitudes and color index values in Table 1 strongly support that LAMOST J052016.79+345651.7 harbors cold circumstellar material.

\section{Analysis of spectroscopic data from LAMOST}

In Sect. 2, we analyze the photometric data of LAMOST J052016.79+345651.7 from the ZTF telescope, as well as the multi-band data from Pan-STARRS, 2MASS, and WISE. In Sect. 3, we analyze 19 LRS spectra of LAMOST J052016.79+345651.7 obtained from LAMOST. LAMOST adopts an innovative active optics technique (Cui et al. 2012), which enables it to combine a large aperture with a wide field of view. From October 24, 2011 to June 30, 2025, the LAMOST DR13 v1.0 released a cumulative total of over 30.8 million spectra, including LRS spectra and medium resolution search (MRS) spectra. LAMOST provides two types of spectra with distinct parameters. Its LRS spectra cover a wavelength range of 3700-9000\,${\AA}$, split into a blue channel (3700-5900\,${\AA}$) and a red channel (5700-9000\,${\AA}$) (Cui et al. 2012), achieving a resolution of 1800 at 5500\,${\AA}$ (Stoughton et al. 2002; Abazajian et al. 2003). In contrast, the MRS spectra focus on 4950-5350\,${\AA}$ for the blue cameras and 6300-6800\,${\AA}$ for the red cameras, delivering a higher resolution of approximately 7500 (Liu et al. 2020). In addition, the MRS spectra also include time-domain data. The parameters of 19 LAMOST LRS spectra for LAMOST J052016.79+345651.7 are shown in Table 2. There is no LAMOST MRS spectrum for LAMOST J052016.79+345651.7. The parameters of $T_{\rm eff}$, surface gravity (log\,$g$), metallicity ([Fe/H]), and radial velocity (Rv) are determined by the LAMOST Stellar Parameter pipeline (LASP, Wu et al. 2014; Luo et al. 2015).

\begin{table}
\begin{center}
\scriptsize
\caption{Parameter table of 19 LAMOST LRS spectra.}
\begin{tabular}{llllllllllll}
\hline
obsdate  &SNR of u,g,r,i,z                         &class           &$T_{\rm eff}$[K]      &log\,$g$           &[Fe/H]             &Rv[km/s]          &H$\alpha$12 \\
\hline
20231009 &\,10.38,\,37.81,\,58.72,\,80.97,\,57.02  &A3              &6224.09 $\pm$ 46.01   &4.178 $\pm$ 0.062  &-0.416 $\pm$ 0.043 &-16.3 $\pm$ 6.04  &-0.11       \\
20231013 &\,13.09,\,37.17,\,55.2,\,77.4,\,53.16    &A0              &6597.55 $\pm$ 45.47   &4.175 $\pm$ 0.061  &-0.491 $\pm$ 0.043 &9.98  $\pm$ 5.9   &1.08        \\
20231014 &\,10.11,\,34.86,\,58.66,\,81.47,\,59.16  &A2              &6431.95 $\pm$ 42.71   &4.018 $\pm$ 0.056  &-0.501 $\pm$ 0.041 &-9.56 $\pm$ 5.31  &-0.18       \\
20231111 &\,10.71,\,24.37,\,24.02,\,30.95,\,19.83  &A0              &                      &                   &                   &                  &-1.38       \\
20231112 &\,9.49,\,24.54,\,24.85,\,30.72,\,19.26   &A0              &                      &                   &                   &                  &-5.46       \\
20231113 &\,6.73,\,21.42,\,21.59,\,26.79,\,17.3    &A2              &                      &                   &                   &                  &-6.4        \\
20231114 &\,8.06,\,21.85,\,17.98,\,17.34,\,10.56   &A2              &                      &                   &                   &                  &-9.11       \\
20231115 &\,7.01,\,19.43,\,18.76,\,22.18,\,13.97   &A2              &6630.53 $\pm$ 152.4   &3.684 $\pm$ 0.246  &-0.032 $\pm$ 0.162 &-52.91$\pm$ 12.62 &-4.14       \\
20231206 &\,7.05,\,16.53,\,16.53,\,21.12,\,14.18   &A0              &6674.34 $\pm$ 160.86  &4.13  $\pm$ 0.264  &-0.551 $\pm$ 0.173 &22.46 $\pm$ 12.26 &-5.11       \\
20231208 &\,8.71,\,19.95,\,19.04,\,25.12,\,16.13   &A0              &                      &                   &                   &                  &-4.33       \\
20240105 &\,8.61,\,21.64,\,21.76,\,28.35,\,17.89   &A0              &6415.14 $\pm$ 121.66  &4.178 $\pm$ 0.194  &-1.465 $\pm$ 0.128 &-20.84$\pm$ 10.73 &-4.57       \\
20240109 &\,7.06,\,21.62,\,23.97,\,31.88,\,20.07   &MWD             &                      &                   &                   &                  &            \\
20240206 &\,5.39,\,15.78,\,29.24,\,53.78,\,40.51   &MWD             &                      &                   &                   &                  &            \\
20241105 &\,7.75,\,37.18,\,58.8,\,78.56,\,57.63    &F2              &6275.32 $\pm$ 41.42   &4.153 $\pm$ 0.056  &-0.5   $\pm$ 0.039 &-3.16 $\pm$ 5.38  &            \\
20241106 &\,5.18,\,31.85,\,57.12,\,77.98,\,56.98   &F3              &6129.14 $\pm$ 40.38   &3.869 $\pm$ 0.056  &-0.477 $\pm$ 0.04  &-11.77$\pm$ 4.7   &            \\
20241205 &\,8.82,\,36.14,\,53.64,\,72.62,\,53.83   &A7              &6595.74 $\pm$ 47.89   &4.318 $\pm$ 0.063  &-0.467 $\pm$ 0.046 &8.01  $\pm$ 6.11  &-1.29       \\
20241224 &\,5.89,\,28.18,\,47.06,\,66.35,\,48.67   &MWD             &                      &                   &                   &                  &            \\
20250103 &\,7.41,\,31.31,\,53,\,74.79,\,54.78      &F0              &6358.68 $\pm$ 42.42   &4.156 $\pm$ 0.06   &-0.54  $\pm$ 0.042 &-14.43$\pm$ 4.87  &            \\
20251021 &\,4.97,\,31.66,\,58.11,\,84.99,\,62.76   &F4              &6220.34 $\pm$ 41.75   &4.076 $\pm$ 0.058  &-0.434 $\pm$ 0.041 &0.2   $\pm$ 4.84  &            \\
\hline
Vizer    &                                         &                &5868.0 $\pm$ 147.0    &4.0567             &                   &                  &            \\
\hline
\end{tabular}
\end{center}
\end{table}

In Table 2, We can see that the signal-to-noise ratio (SNR) in the u-band is relatively low due to atmospheric extinction by the Earth. Considering the five bands u, g, r, i, and z together, the SNR of the first three spectra are relatively high. Therefore, the spectral type of LAMOST J052016.79+345651.7 is most likely A-type. It can be seen from Table 2 that $T_{\rm eff}$ of LAMOST J052016.79+345651.7 is approximately 6200 $\pm$ 400\,K and log\,$g$ is around 4.2. It is metal poor. The Rv can change from -52.91 $\pm$ 12.62\,km/s to 22.46 $\pm$ 12.26\,km/s. The last column is H$\alpha$ line index with band widths of 12\,${\AA}$. These negative values indicate that H$\alpha$ emission line spectra are present.

\begin{figure}
\begin{center}
\includegraphics[width=11.5cm,angle=0]{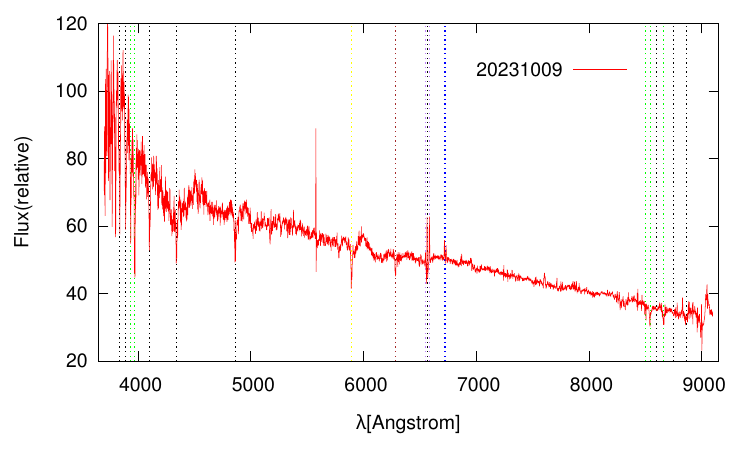}
\end{center}
\caption{LAMOST LRS spectrum for LAMOST J052016.79+345651.7 observed on 2023 October 9. Vertical colored dashed lines mark absorption or emission lines.}
\end{figure}

\begin{table}
\begin{center}
\caption{Wavelength of emission or absorption lines identified in LAMOST LRS spectrum.}
\begin{tabular}{llllllllllll}
\hline
lines                          &wavelength(${\AA}$)              &color        \\
\hline
H$\eta$                        &\,3836.47                        &black        \\
H$\zeta$                       &\,3890.15                        &black        \\
H$\epsilon$                    &\,3971.19                        &black        \\
H$\delta$                      &\,4102.89                        &black        \\
H$\gamma$                      &\,4341.68                        &black        \\
H$\beta$                       &\,4862.68                        &black        \\
H$\alpha$                      &\,6564.61                        &black        \\
Paschen(n=14$\rightarrow$n=3)  &\,8600.80                        &black        \\
Paschen(n=12$\rightarrow$n=3)  &\,8752.91                        &black        \\
Paschen(n=11$\rightarrow$n=3)  &\,8865.37                        &black        \\
Ca II K                        &\,3934.78                        &green        \\
Ca II H                        &\,3969.59                        &green        \\
Ca II                          &\,8500.35,\,8544.44,\,8664.52    &green        \\
N II                           &\,6549.7689,\,6585.1583          &purple       \\
S II                           &\,6718.1642,\,6732.5382          &blue         \\
Na I                           &\,5895.6                         &yellow       \\
Fe I                           &\,6280.8                         &brown        \\
\hline
\end{tabular}
\end{center}
\end{table}

In Figure 3, we show the first spectrum in Table 2. Overall, the spectral features of the Balmer absorption lines are evident. Upon closer inspection, the H$\alpha$ line does exhibit emission line characteristics. Vertical dashed lines of different colors mark the identified absorption or emission lines, as shown in Table 3. The detailed wavelength in Table 3 are from the LAMOST official website or the NIST Atomic Spectra Database (Kramida 2008). We identified the most prominent Balmer lines, Paschen lines, Ca II lines, N II lines, S II lines, Na I line, and Fe I line. The H$\alpha$ line, N II lines, and S II lines are emission spectra. For EW-type binaries, N II and S II emission line spectra are not common. EW-type binary stars have very short orbital periods and are contact binaries with a common envelope. The observation of emission lines indicates the presence of special physical processes.

\begin{figure}
\begin{center}
\includegraphics[width=11.5cm,angle=0]{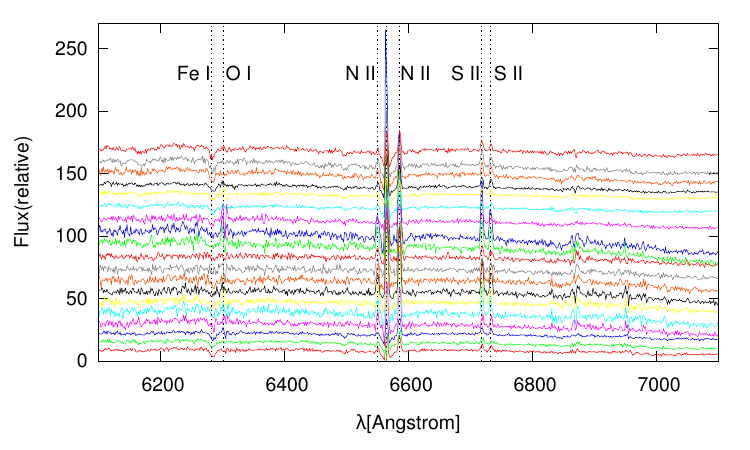}
\end{center}
\caption{Magnified H$\alpha$, N II, and S II emission lines of the 19 spectra.}
\end{figure}

\begin{figure}
\begin{center}
\includegraphics[width=11.5cm,angle=0]{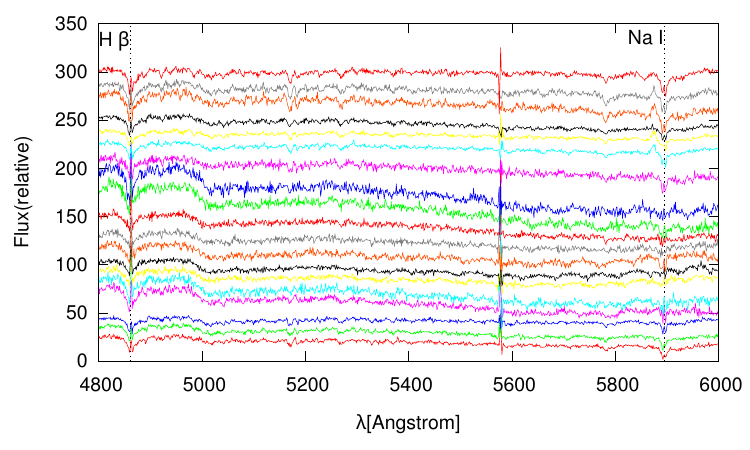}
\end{center}
\caption{Magnified H$\beta$ and Na I absorption lines of the 19 spectra.}
\end{figure}

Figure 4 and 5 show the magnified H$\alpha$, N II, and S II emission lines and H$\beta$, Na I absorption lines of the 19 spectra, respectively. From bottom to top, the vertical order follows increasing time. As shown in Fig. 4, for the 19 spectra covering a duration of two years, the H$\alpha$ emission lines are very prominent, and the N II and S II emission lines are also clearly resolved. The emission intensity of N II (6585.1583 ${\AA}$) is higher than that of N II (6549.7689 ${\AA}$), and the emission intensity of S II (6718.1642 ${\AA}$) is higher than that of S II (6732.5382 ${\AA}$). This intensity relationship implies the presence of optically thin, warm ionized gas with extremely low electron density. Combining the infrared excess in the w4 band of the WISE telescope, as analyzed in Sect. 2, LAMOST J052016.79+345651.7 contains both warm ionized gas and cold dusty material. In Fig. 5, the H$\beta$ emission line is superimposed on the absorption line, and the Na I double yellow lines caused by spin-orbit coupling are clearly visible. Actually, at first glance, the Balmer lines in the overall spectrum in Fig. 3 resemble that of a CV during an outburst (Chen et al. 2026). However, CVs are more often in a quiescent state, and it is impossible for all 19 spectra to be in an outburst phase. Moreover, the light curve directly rules out the possibility of a CV.

\section{Preliminary fitting based on PHOEBE and MESA}

\begin{table}
\begin{center}
\scriptsize
\caption{Best-fit parameters based on PHOEBE. For the models of best1, the reference epoch is T0=58518.90245 and the folding period is P=0.350777\,d, which are consistent with that of Fig. 2. For the models of best2, the reference epoch is T0=58347.4470464 and the folding period is P=0.3507818\,d, which are adopted from chen et al. (2020)}
\begin{tabular}{llllllllllll}
\hline
             &inclination[degree] &q ($M_{2}/M_{1}$) &phase shift     &equivalent radius  &$fillout_{factor}$ &$\chi^{2}_{g}$     &$\chi^{2}_{r}$    &$\chi^{2}_{i}$  \\
\hline
range,step   &35-42,1             &0.80-1.00,0.01    &0.25-0.28,0.01  &1.30-1.60,0.05     &                   &                   &                  &                \\
\hline
best1        &                    &                  &                &                   &                   &                   &                  &                \\
\hline
             &37                  &0.95              &0.26            &1.55               &0.7389             &0.278              &0.214             &0.016           \\
             &37                  &0.96              &0.26            &1.55               &0.7433             &0.278              &0.214             &0.016           \\
             &38                  &0.95              &0.26            &1.50               &0.6160             &0.278              &0.214             &0.016           \\
             &38                  &0.96              &0.26            &1.50               &0.6214             &0.278              &0.214             &0.016           \\
             &38                  &0.97              &0.26            &1.50               &0.6267             &0.278              &0.214             &0.016           \\
             &38                  &1.00              &0.26            &1.50               &0.6418             &0.277              &0.215             &0.016           \\
             &39                  &0.96              &0.26            &1.45               &0.4906             &0.278              &0.214             &0.016           \\
             &39                  &0.99              &0.26            &1.45               &0.5092             &0.278              &0.214             &0.016           \\
             &39                  &1.00              &0.26            &1.45               &0.5151             &0.278              &0.214             &0.016           \\
             &41                  &0.98              &0.26            &1.40               &0.3660             &0.278              &0.214             &0.016           \\
             &41                  &0.99              &0.26            &1.40               &0.3730             &0.278              &0.214             &0.016           \\
             &41                  &1.00              &0.26            &1.40               &0.3799             &0.278              &0.214             &0.016           \\
\hline
range,step   &35-42,1             &0.80-1.00,0.01    &0.49-0.53,0.01  &1.30-1.60,0.05     &                   &                   &                  &                \\
\hline
best2        &                    &                  &                &                   &                   &                   &                  &                \\
\hline
             &37                  &0.82              &0.51            &1.60               &0.7986             &0.218              &0.174             &0.021           \\
             &37                  &0.83              &0.51            &1.60               &0.8036             &0.218              &0.174             &0.021           \\
             &37                  &0.84              &0.51            &1.60               &0.8085             &0.218              &0.174             &0.021           \\
             &38                  &0.84              &0.51            &1.55               &0.6817             &0.218              &0.174             &0.021           \\
             &38                  &0.85              &0.51            &1.55               &0.6877             &0.218              &0.174             &0.021           \\
             &38                  &0.86              &0.51            &1.55               &0.6936             &0.218              &0.174             &0.021           \\
             &40                  &0.80              &0.51            &1.50               &0.5142             &0.217              &0.175             &0.021           \\
             &41                  &0.87              &0.51            &1.45               &0.4259             &0.218              &0.174             &0.021           \\
             &42                  &0.80              &0.51            &1.45               &0.3640             &0.218              &0.174             &0.021           \\
             &42                  &0.81              &0.51            &1.45               &0.3736             &0.218              &0.174             &0.021           \\
             &42                  &0.82              &0.51            &1.45               &0.3829             &0.217              &0.175             &0.021           \\
\hline
\end{tabular}
\end{center}
\end{table}

In Sect. 2 and 3, we can see that LAMOST J052016.79+345651.7 is an EW-type eclipsing binary exhibiting emission line spectra and should also have cold dusty material. LAMOST J052016.79+345651.7 encompasses a wealth of physical processes. In Sect. 4, we employed PHOEBE and MESA to perform preliminary fitting studies on LAMOST J052016.79+345651.7. The two stars share a common envelope, with $T_{\rm eff}$ of both stars taken as 6200\,K. The stellar atmospheres are approximated using blackbody radiation, with a moderate level of limb darkening (a coefficient of 0.5). For the three bands shown in Fig. 2, the observed apparent magnitudes are converted into fluxes and then normalized. The model fluxes are also normalized. The reference epoch for zero phase is taken as T0=58518.90245, and the folding period for the three bands is taken as P=0.350777\,d, the same with that in Fig. 2. The fitting residuals are calculated according to the following equation,
\begin{equation}
\chi^{2} = \sum(F_{obs} - F_{mod})^{2}.
\end{equation}
\noindent In Eq.\,(1), $F_{obs}$ is the normalized observed flux and $F_{mod}$ is the normalized model flux. In PHOEBE, we use four parameters, inclination, mass ratio ($M_{2}/M_{1}$) q, phase shift, and equivalent radius, as the input parameter grid, which correspond to the output parameter $fillout_{factor}$ that reflects the degree of contact of the binary system. The $fillout_{factor}$ ranges from 0 to 1, reflecting the common envelope from just filling the inner Lagrangian point L1 to expanding all the way to the outer Lagrangian point L2. After initial testing and debugging, we set the parameter ranges and step sizes listed in Table 4 and obtained 12 sets of equally best-fit models, as shown for 'best1' in Table 4. The model grid contains 4,704 binary star models. Some parameter combinations lack physical output (e.g., some of those with an equivalent radius of 1.30), resulting in a final output of 4,256 binary star models after fitting. We select the best-fit models by finding the minimum sum of $\chi^{2}$ values across the three bands. The parameters of 12 best-fit models are presented in Table 4, as 'best1'. Except for the phase shift, which can be well constrained, the other three parameters are degenerate, especially for the inclination and the equivalent radius. When the inclination and the equivalent radius are fixed, the fitting results are not sensitive to the mass ratio. A larger equivalent radius corresponds to a larger $fillout_{factor}$, and a larger equivalent radius corresponds to a smaller inclination angle.

\begin{figure}
\begin{center}
\includegraphics[width=11.5cm,angle=0]{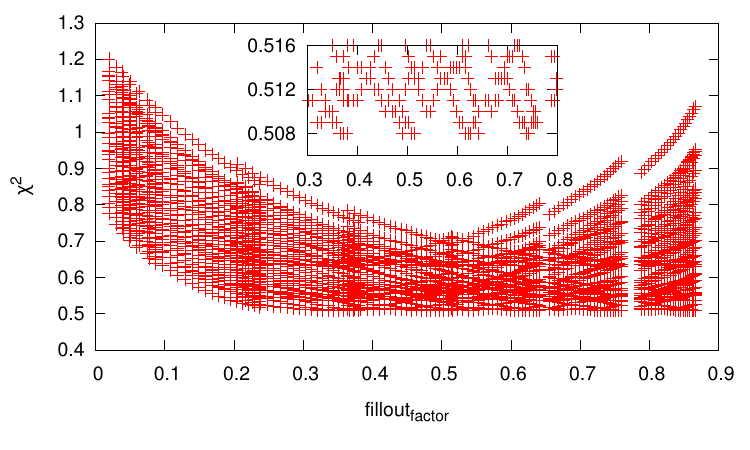}
\end{center}
\caption{Fitting figure of different $fillout_{factor}$ values based on PHOEBE.}
\end{figure}

\begin{figure}
\begin{center}
\includegraphics[width=11.5cm,angle=0]{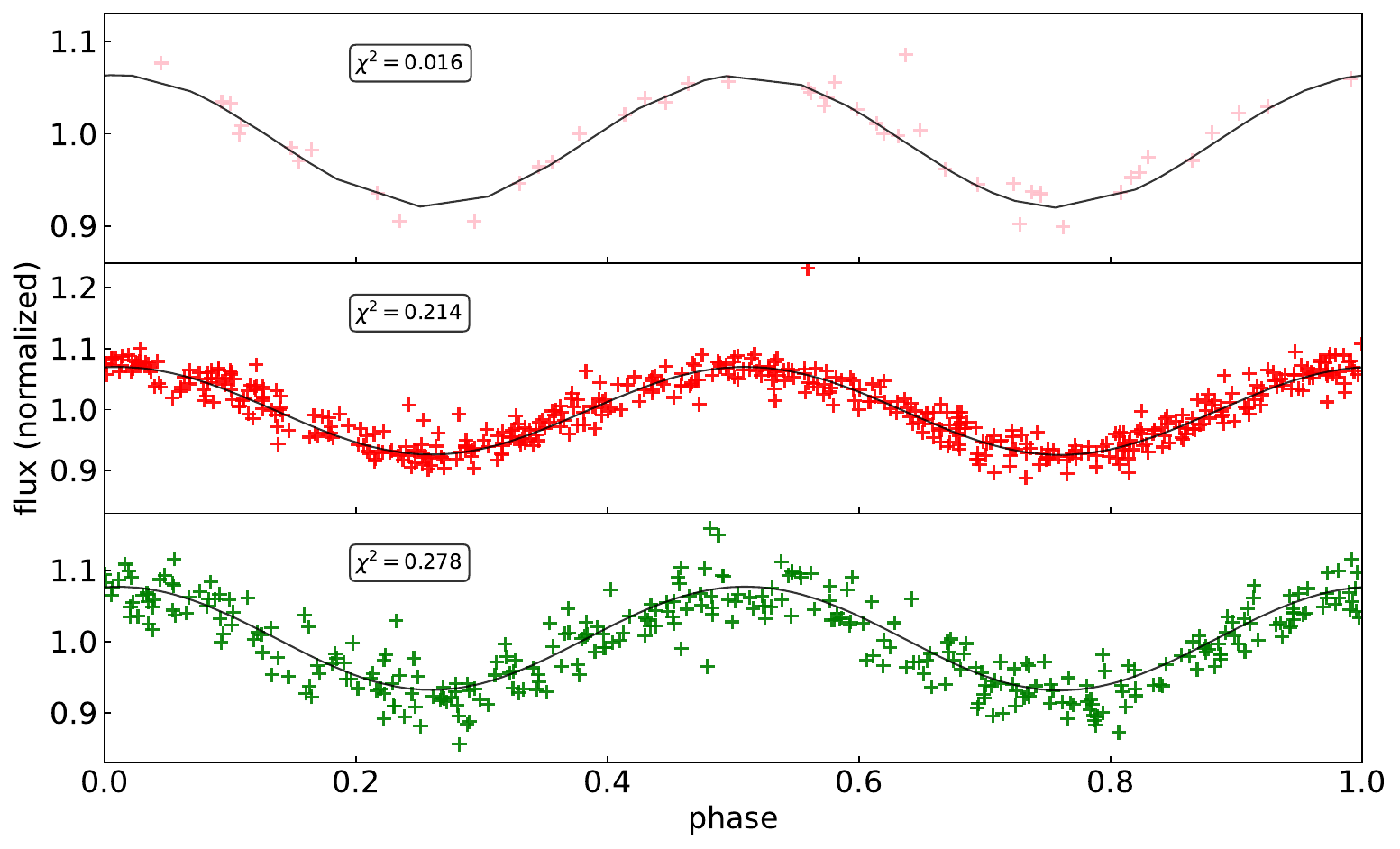}
\end{center}
\caption{Theoretical model fit figure to the ZTF observed phase folded light curves with T0=58518.90245 and P=0.350777\,d based on PHOEBE.}
\end{figure}

\begin{figure}
\begin{center}
\includegraphics[width=11.5cm,angle=0]{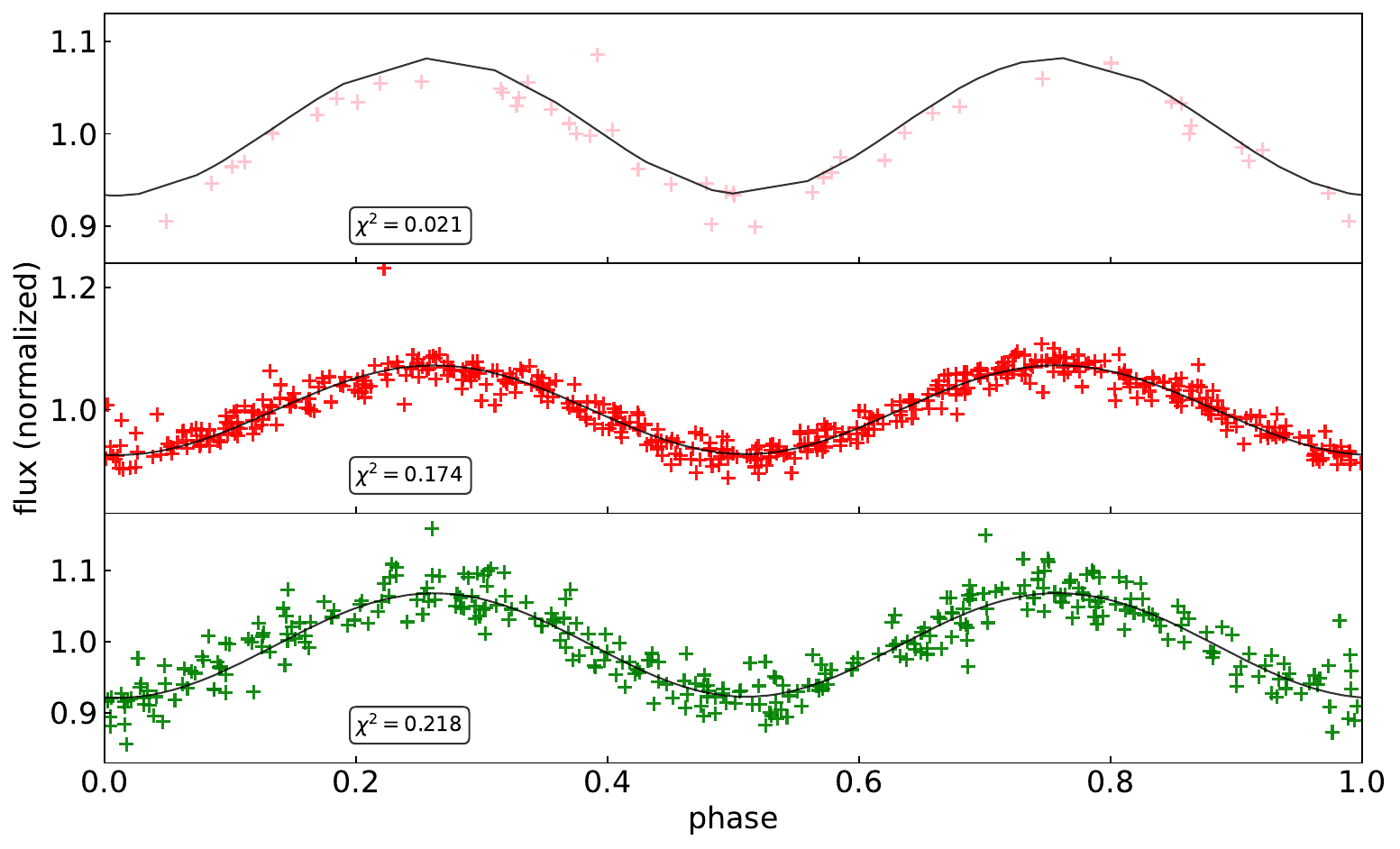}
\end{center}
\caption{Theoretical model fit figure to the ZTF observed phase folded light curves with T0=58347.4470464 and P=0.3507818\,d based on PHOEBE.}
\end{figure}

In Fig. 6, we show the fitting results for different $fillout_{factor}$ values. Due to the degeneracy between inclination and equivalent radius, a certain range of $fillout_{factors}$ (0.37-0.74) can all produce fits with minimal fitting error. The phase folded light curve fits for the three bands of one best-fit model, the one that minimizes the sum of $\chi^{2}$ down to higher decimal precision, are displayed in Fig. 7. We can see that the normalized flux of the model can fit the normalized flux of the observed data points very well. For the 'best2' in Table 4, we additionally calculated the best-fit case with T0=58347.4470464 and P=0.3507818\,d. Only the initial phase of the observational data differs, while the underlying physical laws reflected remain consistent. The model grid contains 5,880 binary star models and a total of 5,320 binary star models are evolved. The sum of $\chi^{2}$ for g, r, and i bands is 0.413 for the 'best2' models. It seems that q is less than 0.90 for the 'best2' models in Table 4. However, if we take the sum of $\chi^{2}$=0.414, then q can include the models with q=1.00. The fitting plot of one best-fit model is shown in Fig. 8, where the phase (phase shift=0.51) is shifted by 0.25 relative to that in Fig. 7 (phase shift=0.26), resulting in an inverted light curve for LAMOST J052016.79+345651.7. This is only a preliminary fit under the blackbody radiation approximation. This star specifically also contains emission line spectra and cold dusty material. In-depth investigation requires input from more complex physical models.

\begin{table}
\begin{center}
\scriptsize
\caption{Model parameters from MESA/binary/test\_suite/evolve\_both\_stars.}
\begin{tabular}{llllllllllll}
\hline
        &period       &velocity             &separation            &radius                &Roche radius           &rl\_relative\_overflow &log\,$g$           &$T_{\rm eff}$    &$fillout_{factor}$  \\
\hline
initial &0.798\,d     &151\,km/s            &4.76\,$R_{\bigodot}$  &1.031\,$R_{\bigodot}$ &1.803\,$R_{\bigodot}$  &-0.41                  &4.47               &5956\,K          &                    \\
end     &0.351\,d     &198\,km/s            &2.75\,$R_{\bigodot}$  &1.241\,$R_{\bigodot}$ &1.042\,$R_{\bigodot}$  &0.19                   &4.31               &6070\,K          &0.6                 \\
\hline
\end{tabular}
\end{center}
\end{table}

Chen et al. (2016) used MESA/binary/test\_suite/star\_plus\_point\_mass to evolve binary stars and simulate the dwarf nova IU Leo. In Sect. 4, we also plan to use MESA/binary/test\_suite/evolve\_both\_stars to evolve binary stars and approximately simulate the evolution of LAMOST J052016.79+345651.7. After repeated attempts, we finally adopted masses of 1.135\,$M_{\bigodot}$ for both components and an initial orbital period of 0.798\,d to simulate the binary evolution. Initially, the two components are detached, and their orbital linear velocity is approximately 151\,km/s. We temporarily ignore the mass loss process and perform a preliminary fit. Considering both angular momentum loss due to magnetic braking (Rappaport, Verbunt \& Joss 1983) and angular momentum loss due to gravitational wave radiation, the orbital period becomes increasingly shorter. In their statistical study of EW-type eclipsing binaries observed by LAMOST, Qian et al. (2017) also proposed that magnetic braking is the dominant mechanism for angular momentum loss in EW-type binaries. When the period reaches 0.351\,d, the orbital linear velocity of the two components is approximately 198\,km/s, and both stars have exceeded their Roche lobes with rl\_relative\_overflow=0.19, sharing a common envelope. The radii of the two components and their Roche radii are 1.241\,$R_{\bigodot}$ and 1.042\,$R_{\bigodot}$, respectively. The Rv from -52.91 $\pm$ 12.62\,km/s to 22.46 $\pm$ 12.26\,km/s in Table 2 most likely trace the kinematics of circumstellar material, not the orbital linear velocity of the two components. The binary has separation, surface gravity, and effective temperature of 2.75\,$R_{\bigodot}$, log\,$g$=4.31, and $T_{\rm eff}$=6070\,K, respectively. It is reasonable that for contact binary stars, the two components are in 'teardrop' shaped ellipsoids, and the sum of their radii is slightly less than the separation between the two stars. The log\,$g$ and $T_{\rm eff}$ are basically consistent with that in Table 2. According to the volume radius calculations of Roche model (Table 6 of Mochnacki 1984),  the $fillout_{factor}$ is approximately 0.6. Adopting the average gravity calculations (Table 14 of Mochnacki 1984), the surface gravity is log\,$g$=4.29, which is basically consistent with the MESA result of log\,$g$=4.31. The parameter $fillout_{factor}$=0.6 is consistent with the best-fit models in Table 4 based on PHOEBE. The corresponding parameters are displayed in Table 5. We use PHOEBE to fit the binary's physical parameters and MESA to fit the binary's evolutionary process. This is only a preliminary fit. Considering mass loss, the future evolutionary endpoint of LAMOST J052016.79+345651.7 should be a blue straggler star.

\section{A Discussion and Conclusions}

Binary star systems contain a wealth of physical laws, and studying them helps to decode interesting physical processes. While searching for magnetic WD (MWD) spectra from LAMOST, we noticed LAMOST J052016.79+345651.7. Its partial emission line spectrum has attracted our attention. Based on the 1D pipeline (Luo et al. 2014), among the 19 LAMOST LRS spectra, three were identified as MWD, while the rest were classified as MS stars. When studying periodic variable sources from the ZTF telescope, Chen et al. (2020) identified LAMOST J052016.79+345651.7 as an EW-type binary system based on its g-band and r-band light curves, with P=0.3507920\,d and 0.3507818\,d (T0=58347.4470464) respectively. A typical EW-type binary consists of two component stars sharing a common envelope and is dominated by absorption lines. LAMOST J052016.79+345651.7 should involve special physical processes.

Using the Period04 program to extract frequencies from the ZTF g, r, and i band data of LAMOST J052016.79+345651.7, the binary orbital periods we obtained are 0.35077\,d, 0.35078\,d, and 0.35086\,d, respectively. Using the average time of the initial observational data from the three bands as reference epoch (T0=58518.90245), the orbital period obtained through phase folding is P=0.350777\,d. Differences in the reference epoch lead to inconsistencies in the phase starting point of the folded light curve, but do not affect the underlying physical laws, as shown in Fig. 7 and 8. The color indices from Pan-STARRS, 2MASS, and WISE, along with the infrared excess in the w4 band, indicate the presence of cold circumstellar material. The 11 smaller radial velocity values measured from the LAMOST spectra in Table 2 also support the presence of circumstellar material. These spectra for LAMOST J052016.79+345651.7 also indicate that $T_{\rm eff}$ is approximately 6200 $\pm$ 400\,K and log\,$g$ is around 4.2. Both the effective temperature and the orbital period of LAMOST J052016.79+345651.7 are typical values for EW-type eclipsing binaries (Li, liu \& Zhu 2020). The 19 LAMOST LRS spectra, covering a duration of two years, all show prominent H$\alpha$ emission lines as well as clear N II and S II emission lines. The intensity relationships of the emission lines indicate the presence of optically thin, warm ionized gas with extremely low electron density. There is no strong evidence for a white dwarf or strong stellar winds. Based on a comprehensive analysis, the emission lines likely originate from shock-driven outflows and mass ejection, while the infrared excess likely comes from dust formed as the outflow cools. Future LAMOST MRS time-domain spectra of LAMOST J052016.79+345651.7 may offer an opportunity to reveal more detailed physical processes.

We performed preliminary modeling of the folded light curves of LAMOST J052016.79+345651.7 using the PHOEBE program, and carried out evolutionary simulation of LAMOST J052016.79+345651.7 from a detached to a contact configuration using the MESA program. Once the reference epoch of the observational data is determined, the phase shift of the fitted model can be uniquely determined, but the inclination and the $fillout_{factor}$ are degenerate. Using MESA to model the evolution of a contact binary, with input masses of both stars set to 1.135\,$M_{\bigodot}$ and an initial period of 0.798\,d, angular momentum loss is driven by magnetic braking and gravitational wave radiation. When the period decreases to 0.351\,d, the resulting parameters roughly match the observed physical parameters of LAMOST J052016.79+345651.7. Considering mass loss, LAMOST J052016.79+345651.7 will evolve to be a blue straggler star. It is a suitable source for studying the evolutionary laws of EW-type binaries with emission lines and circumstellar material.

\section{Acknowledgment}

Guoshoujing Telescope (the Large Sky Area Multi-Object Fiber Spectroscopic Telescope LAMOST) is a National Major Scientific Project built by the Chinese Academy of Sciences. Funding for the project has been provided by the National Development and Reform Commission. LAMOST is operated and managed by the National Astronomical Observatories, Chinese Academy of Sciences. The work is supported by the International Centre of Supernovae, Yunnan Key Laboratory (No. 202302AN36000101) and the Discipline Construction of Physics at Chuxiong Normal University.

\label{lastpage}

\end{document}